# ASFAP impact towards the 1st African Light Source

Gihan Kamel (1 and 2)

*(1)SESAME Synchrotron (Synchrotron-light for Experimental Science and Applications in the Middle East), 19252 Allan, Jordan.*
*(2)Department of Physics, Faculty of Science, Helwan University, Cairo, Egypt.*

---

The concrete vision of having Africa as a leader sharing equivalent responsibilities and deliverables towards the global scientific societies turn out to be more obvious by time. Africa is not an exception when it comes to advanced science and technological grounds. Many challenges do exist and many others are still accumulating such as establishing cutting-edge large scale research infrastructures and institutions, reversing the brain-drain dramatic challenge, addressing local and/or regional concerns (health, environment, water, human heritage), as well as being a vehicle for industrial development and growing economy. In addition to bringing forward the African educational systems, employment status, besides the human capacity building which is alleged to be the backbone of any advanced society. Into the discussion, and besides their strong influence on education and advancing science and technology, as well as, capacity building development, are synchrotron light sources demonstrating the extensive capabilities with numerous techniques supporting a wide range of applications of basic science for instance physics, chemistry and biology, along with applied science aspects including life sciences such as biomedicine, pharmaceuticals and drug design, in addition to agriculture, environment, and air and water pollution, besides materials science and industrial applications, and energy and climate change. Furthermore, comprehensive insights can be identified and documented for cultural heritage and archaeology domains.

***Keywords*** *ASFAP, Light Sources, African synchrotron, Capacity building, economic development*

To address the above multiple challenges and more, a huge demand in the implementation of such infrastructures is evidently viewed. For instance, based on several statistical figures, one of the most important aspects to be also tackled is the gender balance concern. Light sources have also shown to be effective in reducing such a gap as much as possible being an open and flexible environment that is based only on scientific merit and skills. In addition, synchrotron light sources proved to convey a valuable segment of diplomacy — that is based on scientific cooperation ceasing complications across borders. Through them, collaborations were made possible only using the neutral language of science. This in line, can encourage new partnerships on the national and international levels to address mutual demands of scientific and societal challenges, and education and economic development as well. One activity of the African School of Physics (ASP) is the African Conference on Fundamental and Applied Physics (ACP) with its first edition in 2018 at the University of Namibia in Windhoek. In this manuscript, a contribution on the light sources for capacity development and research in Africa that was presented during the second edition of ACP [1], organized on March 7-11, 2022, is reported here.

In an attempt to catch the fast evolving scientific and technical race of light sources around the world, African scientists – through collaborations, agreements and training fellowships – are also in a race with time to set up the first facility ever in the continent. In this contribution, the significant need of such facilities to the African continent – the only continent that is left exclusively without a single synchrotron light source is emphasized (Figure 1). Occasionally, establishing a synchrotron light source goes beyond the financial capacity or even a dedicated initial budget of a single country. Therefore, it represents a real bottleneck for the low economic standing position countries – which is the current case of most African countries. On the other hand, the condition can be also deteriorating as a direct influence by the human capacity deficiency, that yet again, signifies the necessity to reverse the brain-drain issue. Overall, synchrotron light facilities do operate in what can be considered as a democratic mode; that is by facilitating efficacious scientific cooperation to promote peace and understanding between people from different cultures, religions, races, and political systems (Herman Winick).

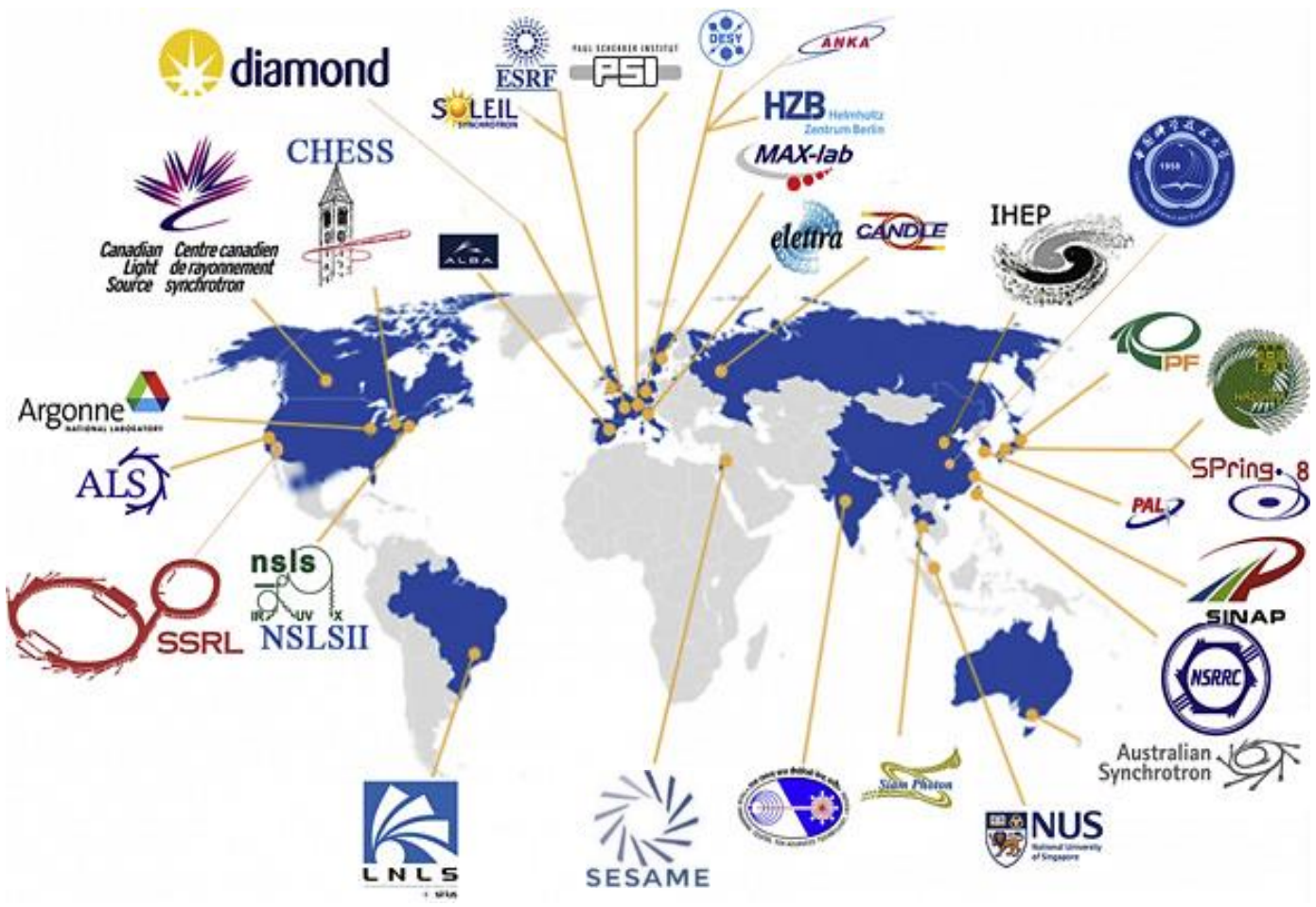

Figure 1: Distribution of synchrotron light sources around the world.

As a general reflection, diverse motivations do embody the aforementioned case, such as the next:

- establishing a world-class and applied research interdisciplinary research laboratories.
- Addressing the many local and regional concerns (for instance; human health, environment, materials and energy, cultural and human heritage, etc.).
- providing a vigorous environment for successful collaborations and allowing the essential space needed for individual career development.
- attracting African diasporas thus drawing back the brain-drain alarm and in the same time resolving the internal brain-drain to other sectors as well, this is the case as the majority may tend to target other fields rather than natural sciences or engineering where the remuneration for jobs in economy for example are much higher than for scientists and with many excellent young scientists choosing such more profitable careers.
- training and preparing graduate students who will no longer need to go abroad to industrialized countries, which implies a minimum of infrastructure and some interesting projects to take place and to be constantly developed in the home country and/or region.
- promoting development of high-tech industry (capacity building).
- decreasing the gender gap on every occasion proves possible.

The ASFAP (African Strategy for Fundamental and Applied Physics [2] basic objective is to develop capacity building in physics education and research. With no exception and as noteworthy as achievements are growing up for other regions, similar scientific and economic challenges persist to be addressed in African continent with the dream that Africa, too, should take its matching identity as a co-leader in the global scientific arena. With this, the requisite of having the ASFAP has turned out to be indispensable for Africa. One of the ASFAP working groups is dedicated to light sources establishment and development. A recent assessment survey launched by the ASFAP Light Sources Working group was released in February. The survey expected to gather a further considerate input from the African scientific community among others on the case of founding an African light source. The subsequent purpose of this questionnaire is to well prepare and establish collaborative research themes and angles. It also aims at preparing Letter of Intents for the Light Sources Working Group within the ASFAP. Herein, some assembled inspirations out of the survey are listed in the following sections.

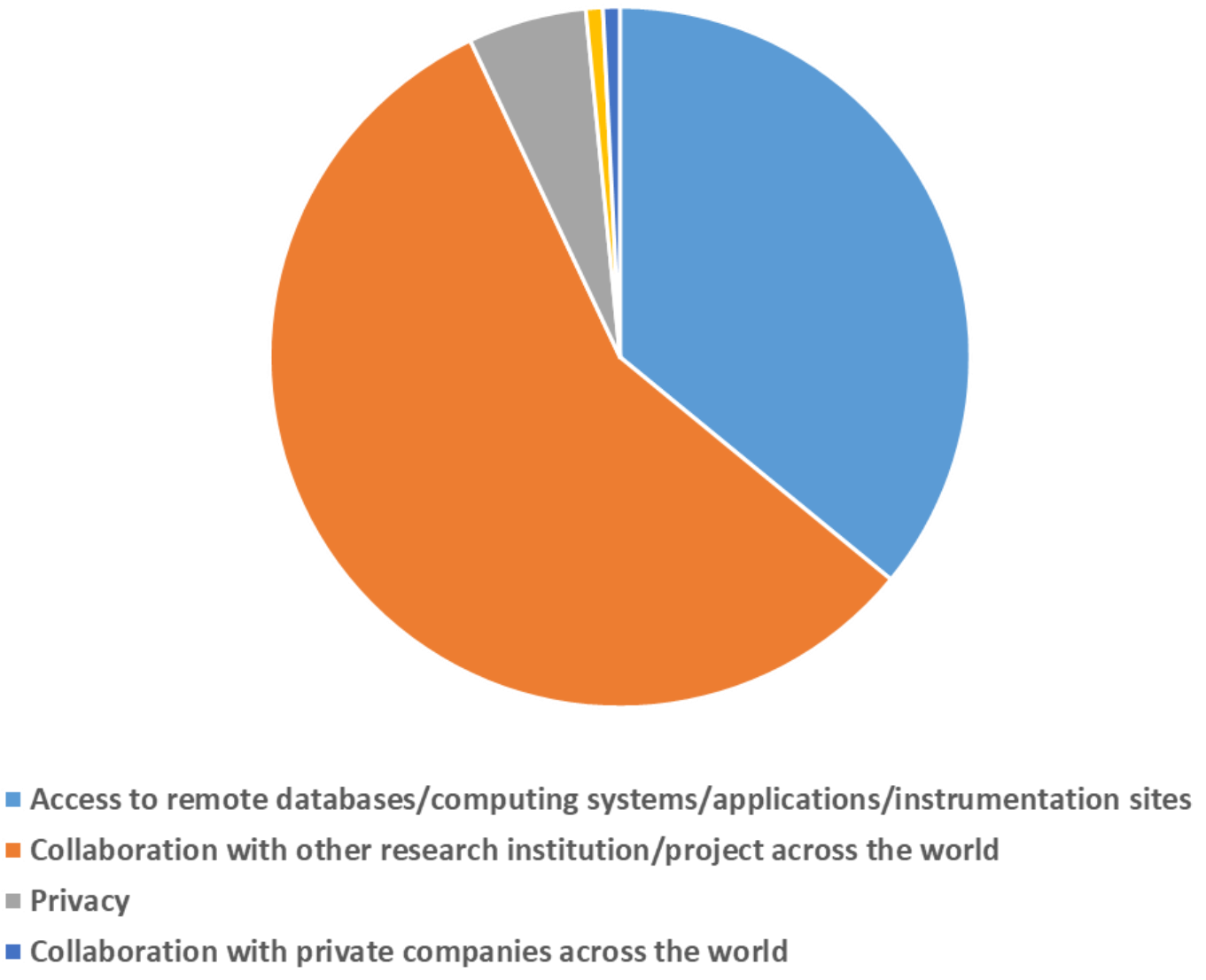

Figure 2: Chief research assisting requirements reported through the ASFAP survey on light sources.

77% of the survey's participants are resident citizens in African countries, while 26% are African diasporas. Participants from nineteen African countries (Nigeria, Morocco, Kenya, Cameron, Senegal, South Africa, Ethiopia, Tunisia, Uganda, Algeria, Ghana, Sudan, Egypt, Ivory Coast, Zambia, Mozambique, Togo, Congo, and Sierra Leon. Participants from 13 non-African countries have also contributed to the survey. Specifically, from USA, India, Pakistan, Italy, Germany, Jordan, UK, France, Malaysia, Peru, Canada, Japan, and Portugal.

Amongst the research interests and scientific activities those were favored by the respondents of the survey came on top the basic and/or applied science, followed by life sciences, materials sciences, cultural heritage and archaeology, accelerators' physics and technology, optical instrumentations, beamlines development, as well as experimental instrumentation and data analysis approaches. A thought-provoking input was also attained by the fact that 76% of the researchers and students opted for current and/or future synchrotron-related interests. Figure 3 shows the required synchrotron techniques, which again, confirms the necessity of establishing such a facility. Moreover, geographical distribution, collaborations with other research institutions, access to remote databases and software, as well as advanced instrumentation, were assigned as higher priorities for research chief requirements. 70% of those who participated showed a previous experience in light sources facilities, while 61% opted for a looked-for employment given the opportunity and depending on qualifications. Besides, 88% opted for their willingness to initiate interactions on different axes of collaboration and assistance with other African groups. 81% marked their need for advanced training regarding the general use of such available infrastructures, with a descending order of financial, technical, and scientific support.

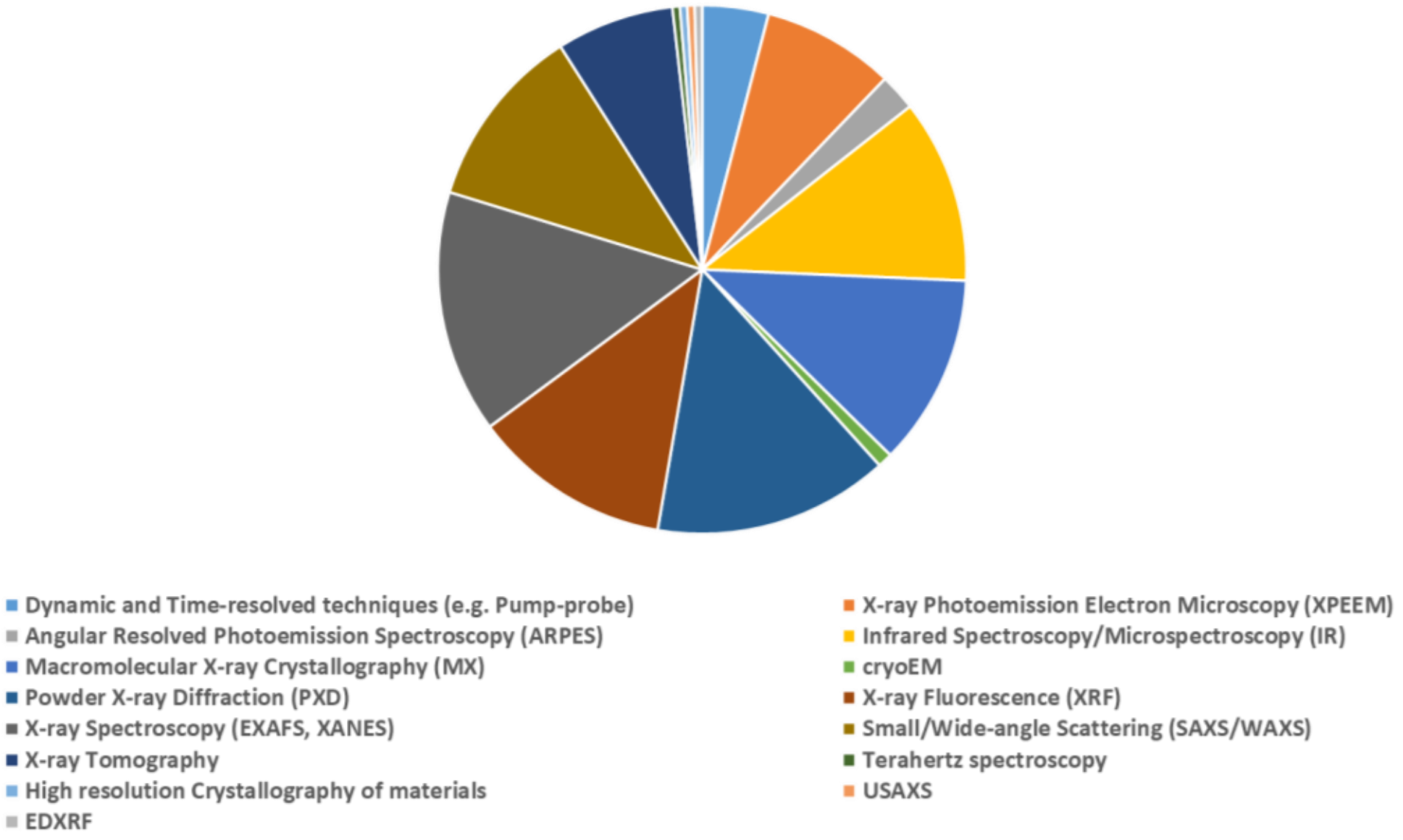

Figure 3: Favorable techniques reported through the ASFAP survey on light sources.

Tackling the expected scientific impacts of light sources have been also brought to the attention of the participants. Some detailed responses are provided subsequently:

- Human health and energy-related materials discovery and development.
- Transfer the know-how among the related counties, and bridging communities through collaborations.
- Profile for African Science, capacity building, local technology, local infrastructure, enhanced networks and participation in international collaborations, more innovation, African wealth.
- A light source facility will support many other research field, providing a framework for research and education centered in Africa. It will also draw the international community and boost the regional economy in providing jobs.
- To make sure that specific problems get attention and not have to depend on exogenous market and policy forces.
- Solving local problems with greater economic output, by means of light sources one can develop solutions and products to raise the balance of trade for Africa.
- Improving major scientific, educational and socio-economic and industrial advances.
- Diversification of the types of research questions posed, particularly in medicine, energy and materials. Escape from European fixation on batteries and fusion.
- With the abundance of mineral resources on the continent of Africa and the huge interest in crystallography, this is a great opportunity to explore our raw materials to create wealth for our people and also education on the interaction of matter with light which will help build the science base of our people.
- To foster scientific and technological excellence; prevent or reverse the brain drain by enabling world-class scientific research; build cultural bridges between diverse societies, as well as education and capacity building.
- Supporting the Pan-African initiative of Africa having its own scientific light source.
- Discovery of novel molecules capable of curing diseases and infections that affect the population.
- Increase number of publications in African countries
- Significant if one would like to keep pace with the global community.

- Light sources technology must be more available and cheaper for all geographical areas in Africa and the world as it provides cutting-edge tools for advancing almost any branch of science.
- Mastering our raw materials and transforming them to get out of poverty.
- Addressing of brain drain and societal issues; Promotion of knowledge base economies

For instance, the status of the human health in Africa represents a huge pillar of scientific research by African scientists and others. Many diseases are there to be investigated and treated. Figure 4 sheds some light on some of the targets to be explored.

There is no doubt that such global research infrastructures do have a strong impact on economy, food security, and disaster management. Moreover, 73% of the participants expected societal impact of light sources in the form of establishing a common culture of knowledge, competitive local industry, entrepreneurship, and capacity building.

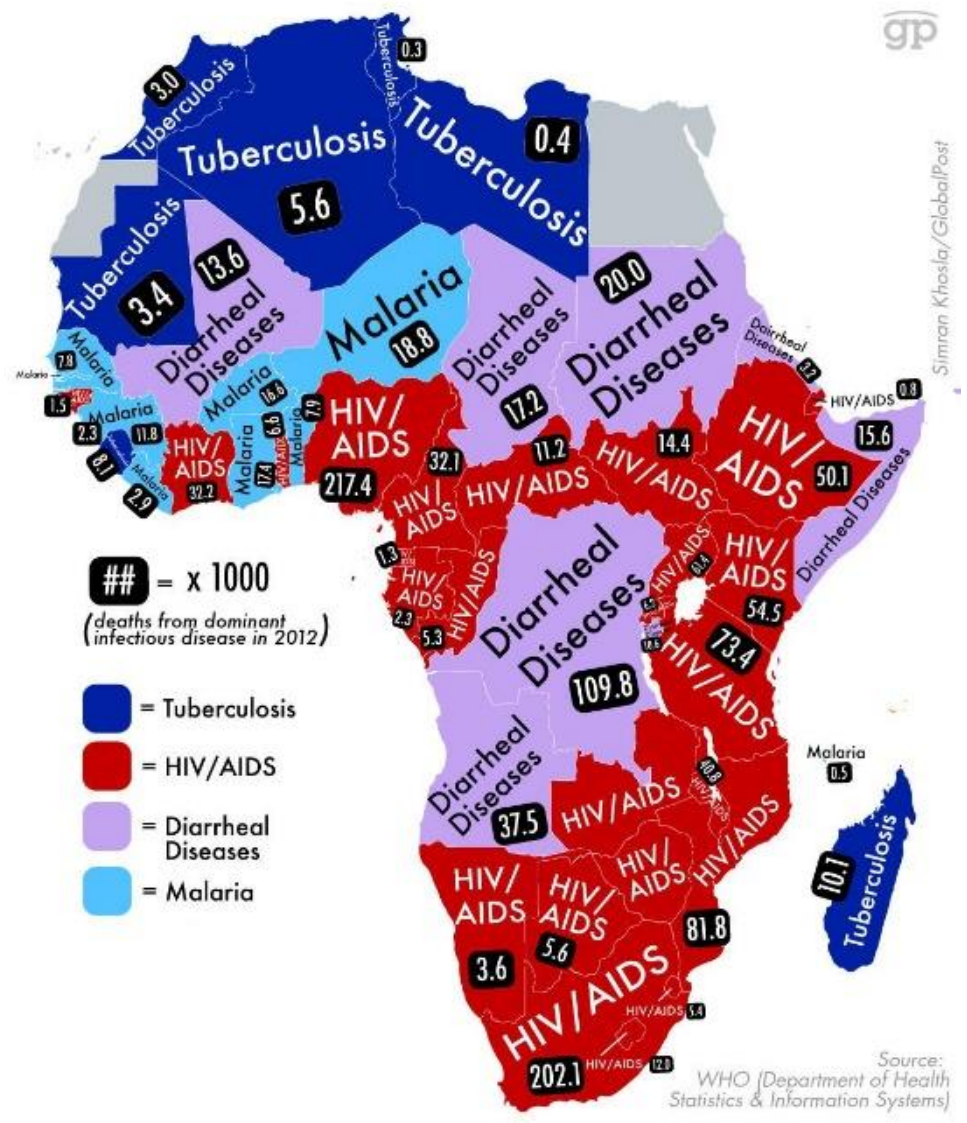

Figure 4: Human health examples of persistent diseases in the African continent.

The participants were also invited to provide their insights on what sort of changes are essential to allow better use of networking facilities. Some collected opinions were as following:

- Prepare short videos highlighting the scientific impact of using synchrotron facilities and addressing what kind of research could be conducted in such facilities. Also, it is important to outreach activities to undergraduate students.
- Scientists everywhere have challenges with stable funding, it is likely more acute in Africa than in the US, EU and Asia.
- Establishment of more local facilities with clustered partnerships (Intra-continental and extra-continental), and sharing equipment available in Africa cross the country and/or within the same through its different institutions.

- Dynamic collaborations to expose the underprivileged institutions.
- Bilateral agreements with African nations by major US and European agencies.

At the same level of importance, and as there is a clear need to have a research infrastructure in Africa specifically and African light source to cope with challenges that Africa is facing, it was also vital to gain some insights from the scientific community on how can African countries join forces to overcome the major challenges to establish its own light source. Below are some of them:

- To start with common infrastructures that can be shared among all communities.
- Cooperation in benchmarking degrees, visas, mobility and exchange funds.
- Involve local industry.
- Develop a concrete strategic vision for a light source facility - Engaging policymakers and the international community to support such a vision.
- In Africa, this might have to be done on region basis via the RECs. Each country, then each REC should develop a major science facility policy in general (as part of STI policies, respectively), and a light source policy in particular. Which can be then developing joint policies given other realities, e.g. transportation routes.
- With the African Union podium and other African institutions in promoting the light source in all African countries and regions.
- By instituting centers of excellence, sharing experiences, equipment and good team work.
- Invest in the science that drives light sources in the rest of the world, e.g. the study of proteins. Plus, showing the necessity of using the light source for research in Africa to solve our local problems such as malaria, famine and technological advancement.
- There must be intense educational system on the research capabilities of light sources and their importance to scientific revolution in Africa.
- Through scientific discoveries and common research activities to tackle preexisting problems and those raised by the side effects of technologies.
- Reach the Critical Mass. Ensure mobility, training, and enrollment of large multi-skilled young scientists through workshops and conferences and funding.
- Collaboration and joint funding to meet the expense of such a huge infrastructure to establish the first African Light Source. African governments can also provide joint funding that involves the private sector. Revenues from minerals and exports should be invested in a light source.
- Establishing better scientific masses within Africa than are groups/projects, much like the commercial and trade blocks that are already existing.
- Top-down and bottom-up organization. Grass roots support is very important. It would be hard to justify "from the top" without strong evidence of current or near-future demand. The multinational aspect of such a project should not be forgotten - it would be a great way for African nations to work closer together. Coming under the umbrella of a Pan-African society such as the AU or perhaps a regional one like SADC, ECOWAS, etc. is another depending on the eventual decisions made. Use the experience gained from SESAME light source on how they are reaching out to new Members to convince them to join.
- Establishment of a regional project, and networking with other light sources.
- Raising awareness among African Heads of State and the African Union on the need to implement their light source for controlled and therefore sustainable development.

Another inquiry of the survey targeted the major obstacles faced when attempting to pursue scientific research in a worldwide light source? Answers were given with the following order.

- Bureaucracy in the facility of destination.
- Lack of funding schemes (travel and mobility, project expenses, etc.).
- Lack of basic and/or preliminary research equipment in own country.
- Lack of mentoring.
- Lack of training opportunities to develop professional skills.
- Scientific merit-related needs.
- Bureaucracy in own country.
- Lack of dedicated manpower.

Based on the above-mentioned information provided within this simple statistical survey, it was also significant to retrieve some informative data on the prospect of the potential cross-disciplinary collaborations and links to light sources user-communities which may be achieved by create multi-folds' links with academia and industrial sectors, as well as, the basic interdisciplinary collaborations. Results showed the following crucial aspects:

- A light source facility can serve communities in various disciplines: materials physics, atomic and molecular physics, biophysics, optics and photonics, Pharmacognosy, etc.
- Materials and Energy systems, biomedical engineering, and plant molecules exploitation.
- Major advances in drug discovery and materials development - crystallographers work with chemists in the extraction and crystallization of potential drug molecules including different vaccine development.
- Agriculture where chemists will synthesize and crystallize fertilizers for crop production, and new techniques to be applied to new fields such as imaging for paleontology, archaeology etc.
- Development of new materials or characterization of newly discovered materials from mines.
- Advancement of society based on communication because it explores wanted field as per the situation.

Light sources are the best example of an open and multidisciplinary research infrastructure. They provide strong opportunities for integration through networking and cost-sharing, and promote multi-disciplinary collaboration with the wider global community, while promoting science diplomacy and peace at large. Moreover, environmental problems, advanced materials, cultural heritage valorization are all complex issue intrinsically involving cross-disciplinary collaboration.

**References:**

[1] Ketevi A. Assamagan, Obinna Abah, Amare Abebe, Stephen Avery, et al., Activity report of the Second African Conference on Fundamental and Applied Physics, ACP2021, arXiv:2204.01882 (2022). doi:https://doi:org/10:48550/arXiv:2204:01882.

[2] Ketevi A. Assamagan, Simon H. Connell, Farida Fassi, Fairouz Malek, Shaaban I. Khalil, et al., The African Strategy for Fundamental and Applied Physics, https: //africanphysicsstrategy:org/ (2021).